\documentclass[conference, onecolumn]{IEEEtran}
\IEEEoverridecommandlockouts
\usepackage{cite}
\usepackage{amsmath,amssymb,amsfonts}
\usepackage{algorithmic}
\usepackage{graphicx}
\usepackage{caption}
\usepackage{subcaption}
\usepackage{comment}
\usepackage{textcomp}
\usepackage{float}
\usepackage{xcolor}
\def\BibTeX{{\rm B\kern-.05em{\sc i\kern-.025em b}\kern-.08em
    T\kern-.1667em\lower.7ex\hbox{E}\kern-.125emX}}

\makeatletter 
\newcommand{\linebreakand}{%
  \end{@IEEEauthorhalign}
  \hfill\mbox{}\par
  \mbox{}\hfill\begin{@IEEEauthorhalign}
}
\makeatother 
    
\begin{document}

\title{Electrification of Clay Calcination: A First Look into Dynamic Modeling and Energy Management for Integration with Sustainable Power Grids \\
\thanks{\textit{EcoClay (Project Agreement No. 64021-7009) is partially funded by the Danish Energy Technology Development and Demonstration Program (EUDP) under the Danish Energy Agency.}}
}

\author{\IEEEauthorblockN{Bruno Laurini}
\IEEEauthorblockA{\textit{DTU Wind} \\
\textit{Technical University of Denmark}\\
Kongens Lyngby, Denmark \\
brulau@dtu.dk}
\and

\IEEEauthorblockN{Nicola Cantisani}
\IEEEauthorblockA{\textit{DTU Compute} \\
\textit{Technical University of Denmark}\\
Kongens Lyngby, Denmark \\
nicca@dtu.dk}
\linebreakand

\IEEEauthorblockN{Wilson R. Leal da Silva}
\IEEEauthorblockA{\textit{Green Innovation} \\
\textit{FLSmidth Cement}\\
Copenhagen, Denmark \\
wld@flsmidth.com}
\and

\IEEEauthorblockN{Yi Zong}
\IEEEauthorblockA{\textit{DTU Wind} \\
\textit{Technical University of Denmark}\\
Kongens Lyngby, Denmark \\
yizo@dtu.dk}
\and

\IEEEauthorblockN{John Bagterp Jørgensen}
\IEEEauthorblockA{\textit{DTU Compute} \\
\textit{Technical University of Denmark}\\
Kongens Lyngby, Denmark \\
jbjo@dtu.dk}
}
\maketitle

\begin{abstract}
\noindent This article explores the electrification in clay calcination, proposing a dynamic model and energy management strategy for the integration of electrified calcination plants into sustainable power grids. A theoretical dynamic modeling of the electrified calcination process is introduced, aiming at outlining temperature profiles and energy usage - thus exploring the feasibility of electrification. The model serves as a tool for optimizing parameters, estimating system behavior, and enabling model-based process control. An innovative energy management model is also presented, ensuring efficient assimilation of electrified calcination plants into the power grid. It encapsulates demand-supply balancing and optimizes renewable energy usage. In essence, we provide an insightful pathway to a more sustainable cement production, underlining the value of renewable energy sources and effective energy management in the context of clay calcination.\\


\end{abstract}

\section{Introduction}

The cement industry is a hard-to-abate sector that contributes to around 8.0\% of the world's CO$_2$ emission \cite{monteiro2017}. In the production process of Ordinary Portland Cement, OPC, (i.e. CEM I \cite{EN197-1-2011}), the emission of CO$_2$ originates from limestone calcination and fuel combustion, with a total of 0.80 $t$ CO$_2$ ⁄ $t$ cement, assuming a coal-fired system. The cement industry has been experimenting with complementary technologies to help reduce its CO$_2$ emissions. The current industrial effort to deliver a more sustainable cement is towards using more Supplementary Cementitious Materials (SCM) in composite cements, while reducing emissions from fossil fuels utilisation via electrification. To that end, calcined clay has gained popularity as SCM due to its worldwide availability, excellent pozzolanic properties, and low cost \cite{Jaskulski2020}. The main industrial solutions for clay calcination rely on the use of fossil fuels. Hence, the replacement of fuel combustion with full electrification can further reduce the CO$_2$ footprint from composite cements. In particular, the use of electricity in a flash calciners requires a electrical hot gas generator to heat the exhaust gas from the preheating cyclones to the desired temperature necessary to calcine clays - usually in the range of 750 - 850$^\circ$C depending on the clay type.

While there exist many positive aspects in electrifying clay calcination process - a comprehensive discussion on technical, economic, and environmental potentials are provided in \cite{brunis2023} and \cite{bruno2024} - the integration of such electrified process with the power grid presents technical challenges. The large increase in electrical demand due to electrification might require investments in both the plant's electrical infrastructure and power transmission system \cite{wei2009}. The latter is dependent on the system's actual power transfer limits, where power demand fluctuations of such large industrial load are to have a considerable impact on the power system's operation. Hence, coordination with system operators and adoption of smart power management strategies are crucial for guaranteeing stability. 

The adoption of demand-side management measures in electrified plants represents an opportunity
for innovation, helping deliver a more sustainable product at a lower operation cost because: a) the use of energy management systems to control dispatchable units (i.e. flexible loads, storage and on-site generation) can help reduce emissions and operational costs, and b) the employment of demand response (DR) strategies by the cement plant can help system operators to handle the intermittence of Renewable Energy Sources (RES), supporting the integration of such resources into the power grid. In view of that, we propose a theoretical dynamic model and energy management strategy for the integration of electrified clay calcination plants into sustainable energy grids.

\section{Theoretical model}
In this section, we describe the mathematical and physical foundation of the cement plant's Energy Management System (EMS) and a dynamic model of the clay calcination process.

\subsection{Energy Management System (EMS)}

An EMS aims to optimize the energy generation and usage inside an energy system. It can be applied to systems of different scales, such as residential and commercial buildings, micro-grids, power plants, and industrial plants. For instance, a residential building can be equipped with photovoltaic modules, lithium-ion batteries, and a HVAC system. In this case, a Building Energy Management System (BEMS) could be used to optimize the power dispatch of flexible technologies for given thermal comfort limits and dynamic electricity contracts (e.g. Time-of-use rates, Real-time pricing) with the utility \cite{ref:BEMS}.

The benefit of using smart energy management strategies is rather minimal in small-scale energy systems. Nonetheless, in an industrial context, they can lead to large costs and CO$_2$ savings. The complexity of EMS increases alongside the scale of the system due to factors such as: distinct processes taking place simultaneously, large critical loads, power distribution complexity, and potential participation in power markets.

From a cement plant outlook, we propose an industrial EMS based on an Optimal Power Flow (OPF) algorithm. Most power distribution networks are based on radial topology, and the power flow within them can be modeled using \textit{Branch Flow Model} (BFM), which focuses on describing the physical relationship between branch variables such as Branch current and power flows.  Assuming a radial distribution topology, the directed graph that represents the network can be described by $G=(N,E)$ . From Ohm's law, the voltage drop between two buses reads:

\begin{equation} \label{eq:BFM1}
    V_i - V_j = z_{ij}I_{ij} \quad \forall (i,j) \in E, 
\end{equation}

where $(i,j)$ denotes a network link - connecting bus $i$ and $j$ - and $E$ represents the set of buses. The branch power flow $S_ij$ definition reads:
\begin{equation} \label{eq:BFM2}
    S_{ij} = V_i I_{ij}^* \quad \forall (i,j) \in E. 
\end{equation}

\begin{figure}[tb]
     \centering
     \begin{subfigure}[b]{0.4\textwidth}
         \centering
        \includegraphics[width=\textwidth]{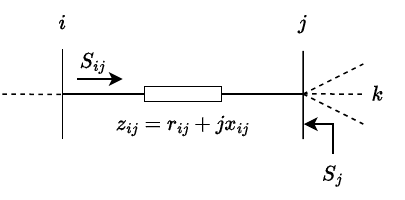}
         \caption{Branch power flow}
         \label{fig:BFM}
     \end{subfigure}
     \hfill
     \begin{subfigure}[b]{0.49\textwidth}
         \centering
         \includegraphics[width=\textwidth]{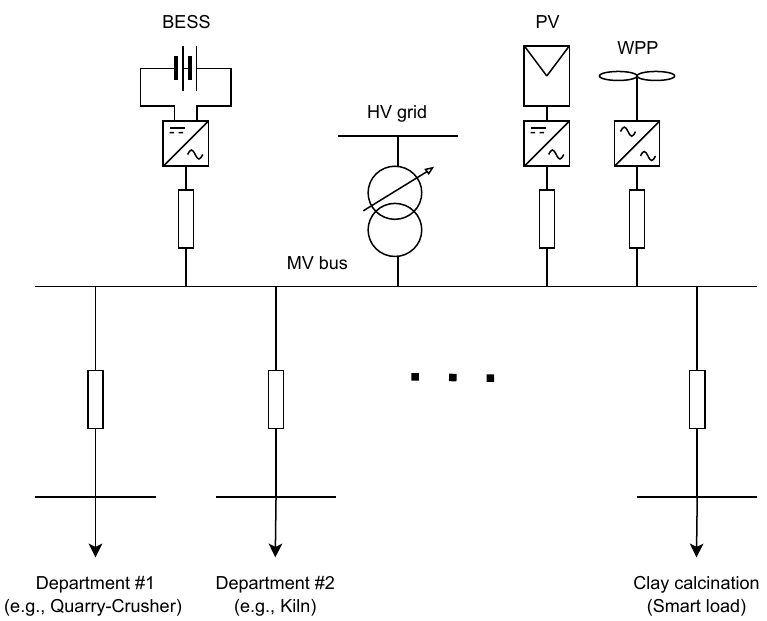}
         \caption{Conceptual topology of the plant's power distribution}
         \label{fig:distribution_net}
     \end{subfigure}
     \hfill
     \caption{Single line diagrams}
        \label{fig:single_line_diagrams}
\end{figure}

Let $N_p(j) \subseteq N $ be the bus set containing the parent of $j$, $N_c(j) \subseteq N $ the bus set of all the children of $j$, and $N_{ss} \subseteq N $ be the set of substation buses; then, the power balance in the branch power flow depicted in Fig. \ref{fig:BFM} can be written as:

\begin{equation} \label{eq:BFM3}
 \sum_{i \in N_{\mathrm{p}}(j)}\left(S_{i j,t}-z_{i j} \left|I_{ij}\right|^2\right) + S_{j,t} = \sum_{k \in N_{\mathrm{c}}(j)} S_{j k,t} \quad \forall j \in N \backslash N_{\mathrm{ss}}.
\end{equation}

Together, (\ref{eq:BFM1}) to (\ref{eq:BFM3}) constitute the branch flow equations and can be used for power flow analysis of a given network. After some manipulation of (\ref{eq:BFM1}) and (\ref{eq:BFM2}), we can arrive in a voltage drop formulation that do not contain complex numbers, only magnitudes, i.e.:

\begin{equation} \label{eq:BFM4}
 U_{j}=U_{i}-2\left(P_{i j} r_{i j}+Q_{i j} x_{i j}\right) + \left(r_{i j}^2+x_{i j}^2\right)l_{i j} \quad \forall(i, j) \in E, 
\end{equation}

where $U$ and $l$ are squared voltage and current magnitudes. The complex power balance in (\ref{eq:BFM3}) can be decomposed into its real and imaginary counterparts by:

\begin{equation}
    \sum_{i \in N_{\mathrm{p}}(j)}\left(P_{i j}-l_{i j} r_{i j}\right)-\sum_{k \in N_{\mathrm{c}}(j)} P_{j k}=P_{j} \quad \forall j \in N \backslash N_{\mathrm{ss}}
\end{equation}
and 

\begin{equation}
    \sum_{i \in N_{\mathrm{p}}(j)}\left(Q_{i j}-l_{i j} x_{i j}\right)-\sum_{k \in N_{\mathrm{c}}(j)} Q_{j k}=Q_{j} \quad \forall j \in N \backslash N_{\mathrm{ss}}
\end{equation}
 
Furthermore, the squared branch current magnitude is related to the voltage drop, and active and reactive power flow through the branch, according to the apparent power definition, i.e.:

\begin{equation}
    l_{ij} = \frac{P_{ij}^2+ Q_{ij}^2}{U_{i}} \quad \forall(i, j) \in E. 
\end{equation}

Along with the power generation/absorption constraints, the aforementioned equations can be cast in a standard optimization form. Such problem represents an OPF model, where a given objective function is minimized given a set of equalities and inequalities constraints. For example, a Distribution System Operator (DSO) could run an OPF aiming at power loss minimization and Conservation of Voltage Reduction (CVR), in a context of high penetration of Distributed Energy Resources (DERs) \cite{ref:dist_OPF}. Our proposed EMS for a cement plant is designed in a similar context, with the main difference being that a DSO cannot control all the loads and generators that are connected to the distribution grid. In the industrial case, the cement plant owns both the network and the loads/generators, i.e. there are many control actions that can be set to achieve the desired objective. 

Figure \ref{fig:distribution_net} shows a conceptual topology of a cement plant's power distribution, where on-site renewable generation (e.g. a PV and a Wind power plant), Battery Energy Storage System (BESS) and the main grid are used to feed both flexible and non-flexible loads. A cement plant is typically divided into functional operating departments according to the different processes. Usually, separate electrical feeder circuits are used to guarantee electrical power supply for each important department depending on their size. The loads of each department require different amounts of active and reactive power that are supplied by the upstream buses.

Moreover, depending on the nature of the process, a load can be flexible or not. In other words, it can modulate its power absorption within a specific range that depends on the process itself. A clay calcination department - containing an electrified clay calcination process - can be a source of power flexibility, especially when coupled with energy storage solutions (e.g. thermal storage). When looking at the plant's power supply part, the main substation represents the dominant source of supply. In this area, Photovoltaic, Wind turbines, and Battery Energy Storage Systems are connected to the AC network through power converters. Such technologies can help increase the plant's power flexibility and reduce both Scope 2 emissions costs and costs of buying power from the utility. Note that, while many industrial plants have fixed electricity contracts with the utility company, there is also an opportunity for the plant to participate in power markets directly, in case it has sufficient power capacity and smart energy management strategies.

\subsection{Dynamic model}

\begin{figure}[tb]
    \centering
    \includegraphics[width=0.6\linewidth]{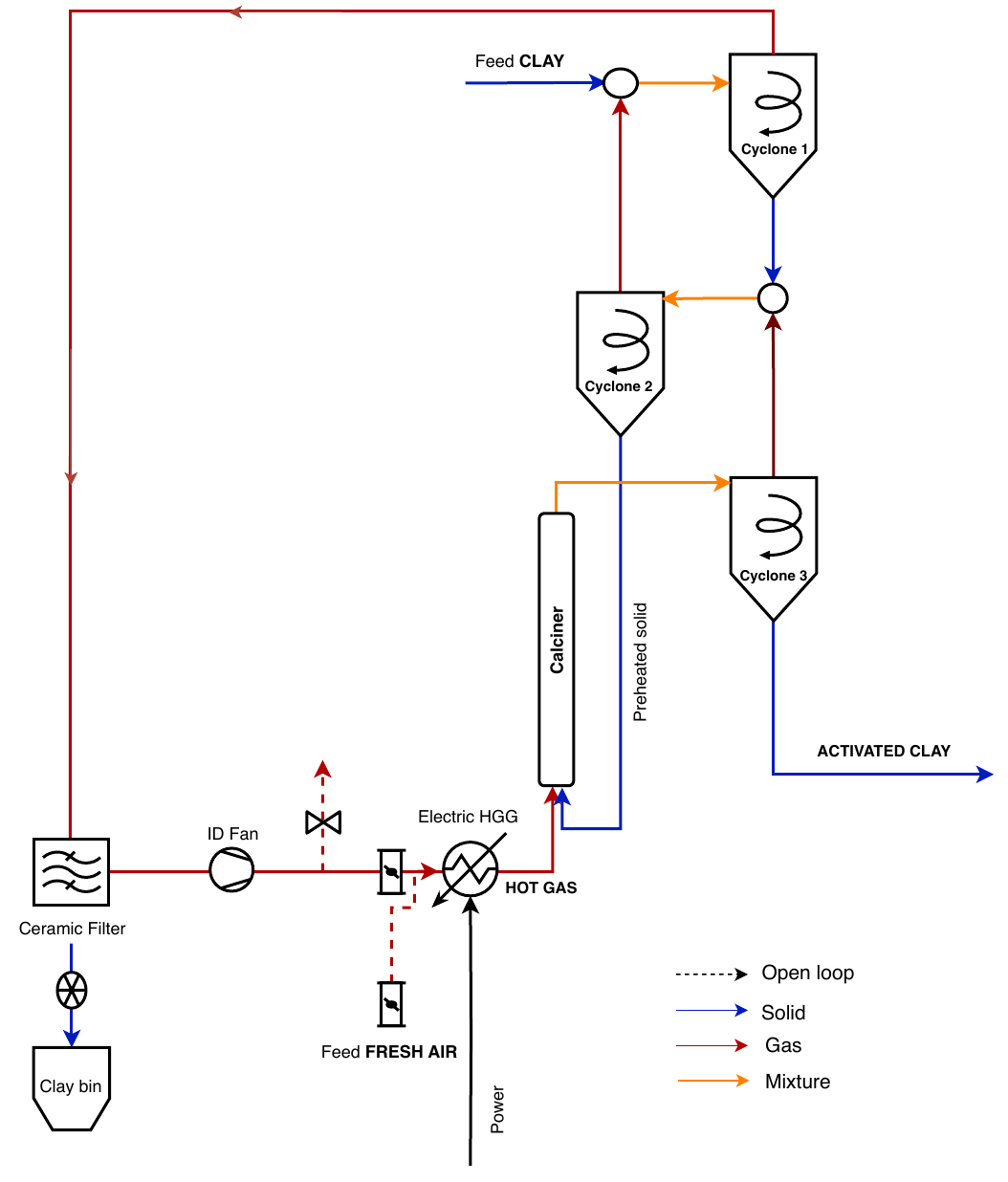}
    \caption{Clay calcination process diagram.}
    \label{fig:ProcessDiagram}
\end{figure}

The establishment of a dynamic model of the clay calcination process is the key to assess and simulate the plant's performance under changing conditions. Such a model is a proxy to evaluate and optimize controllers for the process. Figure \ref{fig:ProcessDiagram} shows the process diagram of the clay calcination in a loop with gas re-circulation. The process starts by introducing ground clay in the first cyclone. The clay is pre-heated by two cyclones before entering the calciner. Next, the calcination reaction is completed in the calciner and the mixture of gas-solid undergoes separation in the last cyclone. Finally, the hot gas is recirculated in a loop. Our approach towards modelling this dynamic process is presented as follows.

The plant-wide dynamic model consists of connected units. The main units are the calciner and cyclones; and this model results in the following system of differential algebraic equations (DAEs):

\begin{subequations}
    \begin{align}
        \dot x (t) = f(x,y,u,d,p), \\
        0 = g(x,y,u,d,p),
    \end{align}
\end{subequations}

where $x$ and $y$ are the differential and algebraic variables. $u,d$ and $p$ inputs, disturbances and model parameters, respectively. The properties and reaction of the materials and substances are evaluated by an independent chemical and thermo-physical model that is shared by all the units. The chemical model comprise stoichiometry and kinetics of the dehydroxylation reaction of clay, i.e.:
\begin{equation}
    \mathrm{Al_2O_2 \cdot 2SiO_2 \cdot 2H_2O (s) \rightarrow } \mathrm{ Al_2O_2 \cdot 2SiO_2 (s)}  \mathrm{ + 2 H_2O (g).}
\end{equation}
This reaction is modeled as a third-order reaction as in \cite{PTACEK201024}. The material's thermo-physical properties are evaluated by building a thermodynamic library, which can compute the material's enthalpy $H$, volume $V$, and internal energy $U$ as a function of temperature $T$, pressure $P$, and number of models $n$ by means of:

\begin{subequations}
    \begin{align}
        V & = V(T,P,n),  \\
        H &= H(T,P,n), \\
        U &= H - PV.
    \end{align}
\end{subequations}

The dynamic behaviour of each unit is modeled by using first principles, i.e. mass and energy balances. Next, the calciner is modeled as plug flow reactor - as fully described in \cite{Cantisani:2024}. The mass and energy balances are expressed as partial differential equations in length $z$ and time $t$, i.e.:

\begin{subequations} \label{eq:calcinerPDE}
    \begin{align}
        \partial_t c &= - \partial_z N + R, \\
        \partial_t \hat u_s & = - \partial_z \Tilde H_s  + \hat J_{sg} + \hat Q_{amb,s}, \\
        \partial_t \hat u_g & = - \partial_z \Tilde H_g - \hat J_{sg} + \hat Q_{amb,g}.
    \end{align}
\end{subequations}

where $c,N,R,\hat u,\Tilde H$ are the concentrations, material fluxes, chemical production rates, volumetric internal energy, and enthalphy flux, respectively. The subscripts $s$ and $g$ indicate solid and gas. $\hat J_{sg}$ and $\hat Q_{amb}$ are the solid to gas heat transfer and heat transfer to the ambient, respectively. 

The velocity of the material stream is modeled explicitly as a function of the pressure drop (by means of the Darcy-Weisbach equation). After spatial discretization of \eqref{eq:calcinerPDE}, the model is completed by adding the algebraic equations:

\begin{subequations} \label{eq:alg_eq}
    \begin{align}
         & V(T_s,P,c_s) + V(T_g,P,c_g) - 1 = 0, \\
         & U(T_{s},P,c_{s}) - \hat u_{s} =0, \\
         & U(T_{g},P,c_{g}) - \hat u_{g} =0.
    \end{align}
\end{subequations} 

The cyclones are modeled as a single cell without spatial discretization. This model consists of mass and energy balances for solid and gas, formulated as DAEs in the same fashion as the calciner. The material fluxes depend on the cyclone separation efficiency, which is modeled explicitly as in \cite{Muschelknautz:1993}. The velocities at the inlet and outlet are modeled as a function of the pressure drop between the input and output streams.

Other minor process units (e.g. electric hot gas generator - EHGG, circulating fan and particle filter) are modeled using static relationships and incorporated in the model as algebraic equations. In addition, the connection between the units is realised by lumping the flow resistance in the connecting tube as a pressure node \cite{Hansen:1998}. The flow rate $F$ between 2 units (1 and 2) reads:
\begin{equation}
    F = C \sqrt{\frac{(P_1 - P_2)(P_1+P_2)}{T_1}},
\end{equation}
where $C$ represents the flow resistance.

The inputs $u$ (variables) of the plant-wide model are the clay feed, pressure rise after the circulating fan, and fresh air intake. The disturbances $d$ are the electrical power to the hot gas generator and the ambient temperature. Notice that the model may be easily extended to a system of stochastic differential algebraic equations to account for different sources of uncertainty and model/plant mismatch. Nonetheless, such extension is beyond the scope of this publication.

\section{Model integration}

The integration of the EMS and Dynamic calcination models provides a link between the electricity supply of the cement plant and the thermodynamics of the clay calcination process. Such a link allows for emulating the physical constraints of both systems and investigating how the plant can operate close to optimal conditions from costs, emissions, and technical standpoint.  While both models are linked through the power absorbed by the EHGG, they have different temporal/spatial scales, following a similar modelling approach to the hierarchical algorithm for integrated scheduling and control proposed in \cite{john2017}. The EMS is based on a systemic approach, where the focus is on modeling the electrical interaction of different processes through the power distribution network and scheduling dispatchable units.

The scheduling problem is directly related to daily production plan of the cement plant and to wholesale electricity markets. Therefore, a model with relatively large time-horizon (e.g. 24 hours) and low frequency (e.g. 1 hour) is used. Conversely, the dynamic model of the clay calcination plant enables a detailed simulation of the process on the temporal scale of seconds, capturing transient states and detailed process dynamics.  

The dynamic model keeps track of crucial variables such as temperature in the calciner, calcined clay at the outlet, calcination degree in every step of the process, etc. A local plant controller is implemented to control the process. As mention in Section II.A, the EMS aims to optimize the energy utilization in the cement plant considering techno-economic and emission-wise aspects, i.e. the objective function reads:

\begin{subequations} \label{eq:EMS_opt}
    \begin{align}
        \min \quad & \phi = \phi_c + \phi_{CO_2} + \phi_{u} - \phi_{cc} \\
        \text{s.t.} \quad & \text{Network constraints}, \\
         & \text{Operational/process constraints}.
    \end{align}
\end{subequations}

The terms in the objective function are operational cost including electricity purchase ($\phi_c$), CO$_2$ emissions ($\phi_{CO_2}$), voltage deviation ($\phi_u$), and calcined clay production ($\phi_{cc}$).

\begin{figure} 
    \centering
\includegraphics[width=0.8\linewidth]{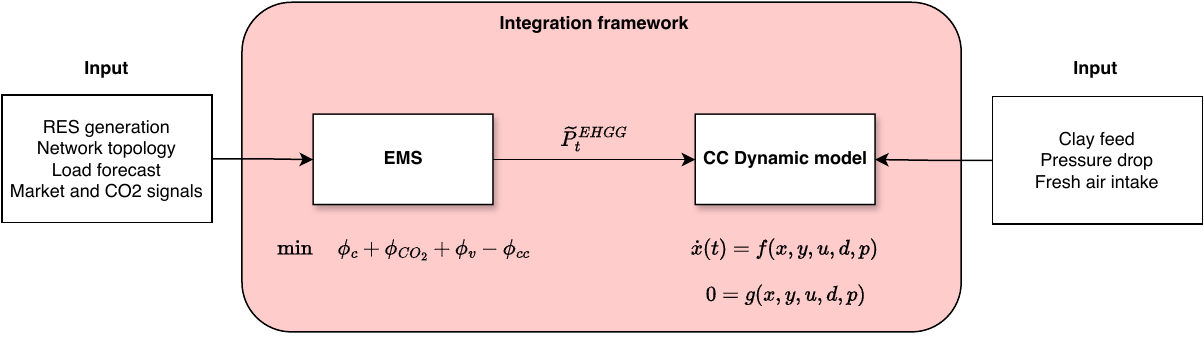}
    \caption{Integration framework of the EMS and dynamic calcination model.}
    \label{fig:integration_framework}
\end{figure}

Solving the optimization problem \eqref{eq:EMS_opt} will provide scheduling decisions for the chosen time horizon. These may include power set-points for generation units and storage, or the operational status of voltage controlling devices. The use of different objective terms characterizes a multi-objective optimization and allows for the EMS to make scheduling decisions based on a combination of different objectives. Weights may need to be used in the objective function to tune the performance of the optimizer. Amongst the decision variables, $\widetilde P^{EHGG}$ represents the power set-point sent to the clay calcination plant controller.

Note that the EMS sets the active power consumed by the EHGG based on a higher-level model: as such, it does not capture the dynamics of the clay calcination process. The actual power used by the EHGG in real-time will actually depend on the process dynamics and the plant controller. Fig. \ref{fig:integration_framework} shows the proposed model integration framework.


\section{Conclusion}

This paper presents a theoretical methodology aimed at optimizing energy management in cement plants, with a focus on an electrified clay calcination process. We propose an EMS model alongside a dynamic model for the clay calcination. Also, we discuss an integration framework to link the power management of the plant with the electricity requirements of the electrified calcination process. This framework is designed to reduce operational costs associated with electricity procurement, while mitigating indirect carbon dioxide emissions from the power grid. By addressing technical and production-related constraints, our approach aims to achieve these objectives while ensuring the fulfillment of production targets and technical requirements. Future work should be carried out to extend the EMS and dynamic models including a market bidding strategy that encompasses participation in Day-ahead and balancing electricity markets and advanced control strategies based on model predictive control.

\section{Acknowledgments}

The EcoClay project (Project Agreement No. 64021-7009) - which is partially funded by the Danish Energy Technology Development and Demonstration Program (EUDP) under the Danish Energy Agency.

\bibliography{document.bib} 
\bibliographystyle{ieeetr}

\vspace{12pt}

\end{document}